\documentclass[preprint,12p]{revtex4}

\usepackage[brazil, english]{babel}
\usepackage[utf8]{inputenc}
\usepackage{graphicx}

\usepackage{caption}
\usepackage{subcaption}
\usepackage{mathtools}
\usepackage{epsfig}
\usepackage{amssymb}
\usepackage{amsmath} 
\usepackage[unicode=true,bookmarks=false,breaklinks=false,pdfborder={0 0 1},colorlinks=true]
 {hyperref}
\hypersetup{
 citecolor=blue,linkcolor=blue,urlcolor=blue}

\graphicspath{%
    {converted_graphics/}
    {/}
}

\begin{document}

\title{Thermodynamics and remnants of Kiselev black holes in Rainbow gravity}
\author{P. H. Morais$^{1}$, G. V. Silva$^{1}$, J. P. Morais Graça$^{1,2}$ and V. B. Bezerra$^{1}$}
\email[Eletronic address: ]{jpmorais@gmail.com; phm@academico.ufpb.br}

\affiliation{$^1$Departamento de F\'isica, Universidade Federal da Para\'iba, \\Caixa Postal 5008, 58059-900, Jo\~ao Pessoa, PB, Brazil 
\\ $^2$Instituto de F\'{\i}sica, Universidade Federal do Rio de Janeiro, 21.941-972 - Rio de Janeiro-RJ - Brazil}

\begin{abstract}
We do a detailed study of the thermodynamics of black holes in Rainbow gravity in the framework of the Kiselev metric, taking into account different cosmic fluids which can mimic several kinds of dark energy scenarios. A discussion related to the appearance of black hole remnants is presented and, unlike other studies on the same topic, we found no evidence of the appearance of remnants in Rainbow gravity, except in peculiar cases.
\end{abstract}


\maketitle

\section{Introduction}
The special theory of relativity is grounded in two basic pillars, that physics should be the same for all inertial frames and that the speed of light is constant. These pillars were necessary to reconcile kinematics with electrodynamics and to explain the Michelson-Morley experiment. Its development by Albert Einstein in 1905 has guided theoretical physics since then, leading to a generalization to non-inertial frames ten years later, named the general theory of relativity.

Nowadays, the idea that special relativity should be modified to incorporate new theoretical and observational evidence is a current area of research. From the theoretical point of view, the major problem is the absence of a  well-defined quantum theory for gravity, and from an observational perspective, the question of ultra energetic cosmic rays has posed a challenge on the dispersion relation established by special relativity \cite{AmelinoCamelia:2000zs,AmelinoCamelia:2000ev}.

A common feature of some quantum theories of gravity, such as string theory \cite{Polchinski:1996na, Forste:2001ah} and loop quantum gravity \cite{Rovelli:1997yv,Carlip:2001wq}, is the modification of the dispersion relation, i.e., the relation between momentum and energy of a particle (in momentum space). Among these theories with modified dispersion relation (MDR) one can cite the so-called double special relativity (DSR) \cite{AmelinoCamelia:2000ge,AmelinoCamelia:2002wr,AmelinoCamelia:2000mn,Magueijo:2002am}, which can be considered an extension to special relativity with not just one, but two constants of nature: the speed of light and the Planck energy scale. Since energy is not an invariant for Lorentz transformations, it is expected that DSR leads to new (and non-linear) coordinate transformations. 

To circumvent the fact that we do not know, a priory, the form of such transformations, Magueijo and Smolin proposed a generalization to include spacetime dynamics in the theory with no reference to position space \cite{Magueijo:2002xx}. From a general formula of the new dispersion relation, 

\begin{equation}
    E^2 f^2\left(\frac{E}{E_p}\right) - p^2 g^2\left(\frac{E}{E_p}\right) = m^2,
    \label{mdr}
\end{equation}

\noindent
where $E_p$ is the Planck energy scale, $f$ and $g$ are phenomenological functions of energy of the particle, dubbed as Rainbow functions, and $m$ is the mass of the particle, the gravitational field equations must now be given by

\begin{equation}
    G_{\mu\nu} \left(\frac{E}{E_p} \right) = 8 \pi G T_{\mu\nu} \left(\frac{E}{E_p} \right). 
\end{equation}

This means that the metric of the spacetime as felt by a test particle with energy $E$ must depend on its energy. For consistency, it is required that in the limit of low-energy the Rainbow functions should be $\text{lim}_{E_p \rightarrow 0} f(E/E_p), g(E/E_p) = 1$. This is the origin of the terminology \textit{Rainbow gravity}, the metric is no longer the same for all test particles but an one parameter family of metrics with the energy of the test particle as the parameter.

As stated above, at the limit of a low-energy scenario, Rainbow gravity should reduce to general relativity, but in a regime of strong gravity, like in the vicinity of black holes, we should expect discrepancies between theories. In this scenario, with the creation of virtual pair of particles and the subsequent emission of Hawking radiation \cite{Hawking:1974sw}, one can study the evolution and fate of black holes. This study is important for several reasons, among them we can mention the fact that, if Hawking radiation stops at some point before total evaporation, the appearance of a remnant black hole \cite{Giddings:1992hh} can help solve two major problems in physics today: the information paradox \cite{Chen:2014jwq} and, perhaps, an origin for dark matter \cite{Chen:2002tu}.

The main goal of this paper is to study black hole thermodynamics in rainbow gravity \cite{Ling:2005bp, Mu:2015qna} in the presence of a cosmic fluid that can mimic several kinds of dark energy. The reason for incorporating a cosmic fluid is because it is as relevant to cosmology today as it was in the early universe. To mimic such a scenario, we will adopt, in this paper, the Kiselev's black hole \cite{Kiselev:2002dx} as the background.

The appearance of black hole remnants in Rainbow gravity has been debated in the literature in recent years, with conflicting opinions among the authors \cite{Ali:2014xqa,Gim:2014ira,Kiselev:2002dx}. With or without the incorporation of a cosmic fluid, we will show that we have found no evidence that Rainbow gravity, by itself, will lead to their appearance, except for a very peculiar case.

This article is divided as follows. In section two, we will present the formalism of black hole thermodynamics in Rainbow gravity, as well as we will discuss black hole thermodynamics in special cases, such as in the absence of the cosmic fluid in Rainbow gravity, and in the presence of the cosmic fluid in general relativity. The goal is that, in the next section, when all the ingredients are brought together, the reader is already more familiar with the results we are looking for. In section three, we will study thermodynamics and black hole remnants in the full picture, in Rainbow gravity with the presence of a cosmic fluid. Finally, in section four, we will present our conclusions.

\section{Black hole thermodynamics in rainbow gravity}

One of the main issues in MDR theories is how to move from momentum space to coordinate space and then to the metric of spacetime. The most accepted proposal so far is that of Magueijo and Smolin \cite{Magueijo:2002xx}, where the metric can be written in terms of the functions $f(E/E_p)$ and $g(E/E_p)$ as $ds^2 = - (A/f) dt^2  + \sum_{i=1}^{3} (B_i/g) (dx^i)^2$, where $A$ and $B$ are functions of the coordinates. 

Using this prescription, a non-rotating black hole in Kiselev spacetime can be described by the following spherically symmetric metric  

\begin{equation}
    ds^2 = - \left(1- \frac{2 GM}{r} - \frac{c}{r^{3\omega + 1}}\right) \frac{1}{f^2}dt^2 + \left(1-  \frac{2 GM}{r} - \frac{c}{r^{3\omega + 1}}\right)^{-1}\frac{1}{g^2}dr^2 + \frac{r^2}{g^2} d\Omega,
    \label{metric}
\end{equation}

\noindent
where $d\Omega = d\theta^2 + \text{sin}\theta \; d\phi^2$, $c$ is a constant and $\omega$ is a parameter related to the equation of state of the surrounding matter. Considering a black hole in a space with a cosmological constant, then $\omega = -1$. Other interesting scenarios are $\omega = -2/3$ and $\omega = -4/3$, which mimic, respectively, a kind of quintessence and phantom matter.

In this paper, we will follow the traditional recipe to calculate the Hawking temperature of the black hole,

\begin{equation}
    T_H = \frac{\kappa}{2\pi},
\end{equation}

\noindent
where $\kappa$ is the surface gravity. For a spherically symmetric black hole, it is given by

\begin{equation}
    \kappa = \lim_{r \rightarrow r_+} \sqrt{- \frac{1}{4} g^{tt} g^{rr} \left(\frac{\partial g_{tt}}{\partial r}\right)^2},
\end{equation}

\noindent
where $r_+$ is the horizon radius. The temperature will depend on the parameters of the theory, and concerning the Rainbow functions it will be given by

\begin{equation}
    T_H = \frac{g(E/E_p)}{f(E/E_p)} \Tilde{T_H},
    \label{Temp}
\end{equation}

\noindent
where $\Tilde{T_H}$ is the temperature of the equivalent spacetime in general relativity. This means that, when $f(E/E_p) = g(E/E_p)$, the Hawking temperature in both theories will be the same. As mentioned in the introduction, the Rainbow functions must be chosen through some theoretical or phenomenological basis, and we will rely on advances in loop quantum gravity to choose

\begin{equation}
    f(E/E_p) = 1, \hspace{20pt} g(E/E_p) = \sqrt{1-\eta\left(\frac{E}{E_p}\right)^n},
\end{equation}

\noindent
where $\eta$ is a dimensionless parameter.

The mass of the black hole is given by the parameter $M$ of the metric ($\ref{metric}$) when the radius is the radius of the horizon, 

\begin{equation}
    M = \frac{r_+}{2G} - \frac{c}{2 G r_+^{3\omega}}.
    \label{eqMass}
\end{equation}

\noindent
Entropy can then be calculated using the first law of thermodynamics, $dU = T dS$, considering the mass of the black hole equivalent to the internal energy $U(S)$. In this paper we will disregard any kind of work, as our goal is the study the behaviour of temperature, specific heat, entropy and decrease in mass as a function of the mass/radius of the black hole. 

We have (almost) all the necessary ingredients to study black hole thermodynamics and remnants, but before we proceed let us make some digressions.
 
\subsection{Black holes in Rainbow gravity}

\begin{figure}[!t]
     \centering
     \begin{subfigure}[b]{0.45\textwidth}
         \centering
         \includegraphics[width=1.25 \textwidth]{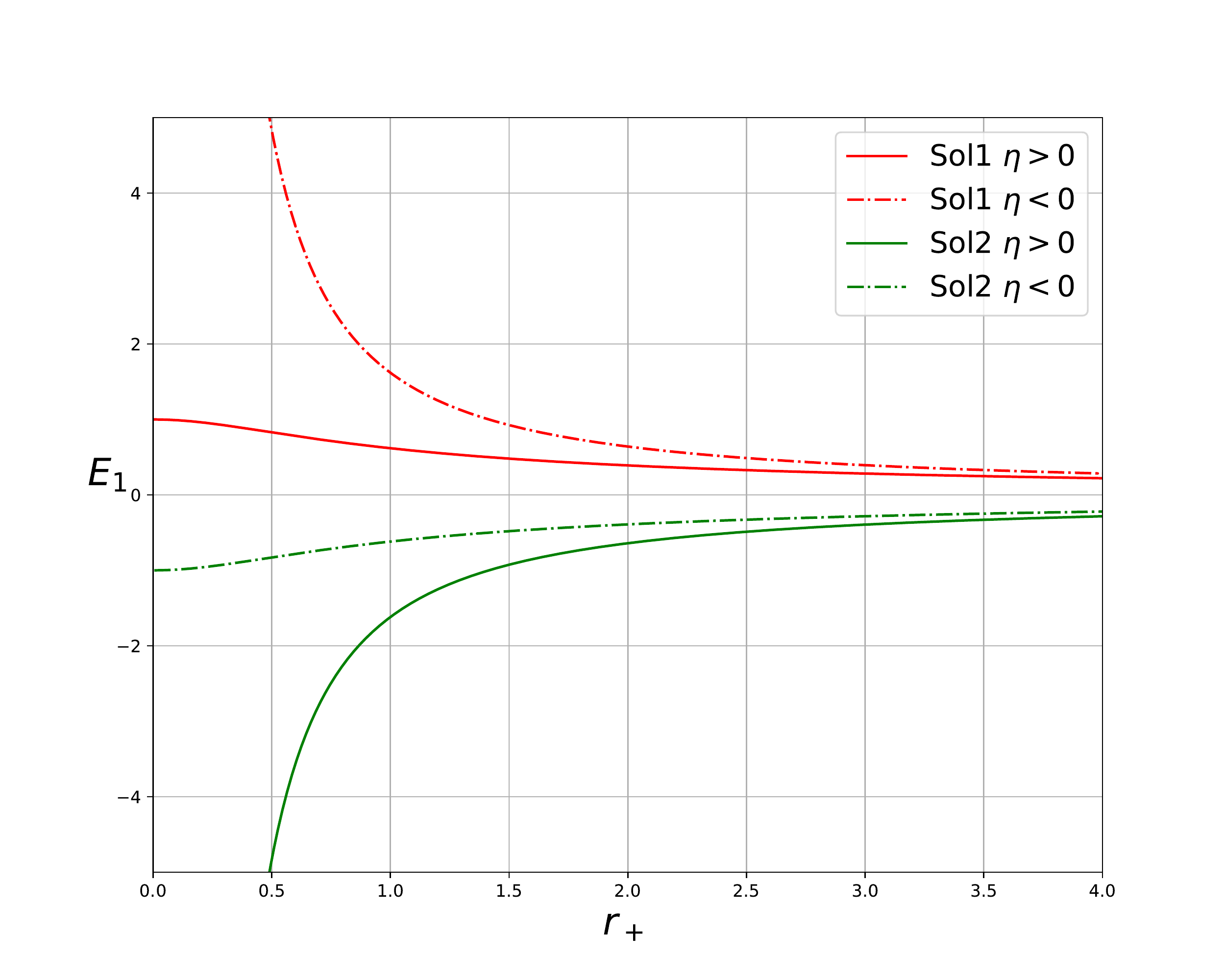}
         \caption{$n=1, |\eta| = 0.9$}
     \end{subfigure}
     \hfill
     \begin{subfigure}[b]{0.45\textwidth}
         \centering
         \includegraphics[width=1.25 \textwidth]{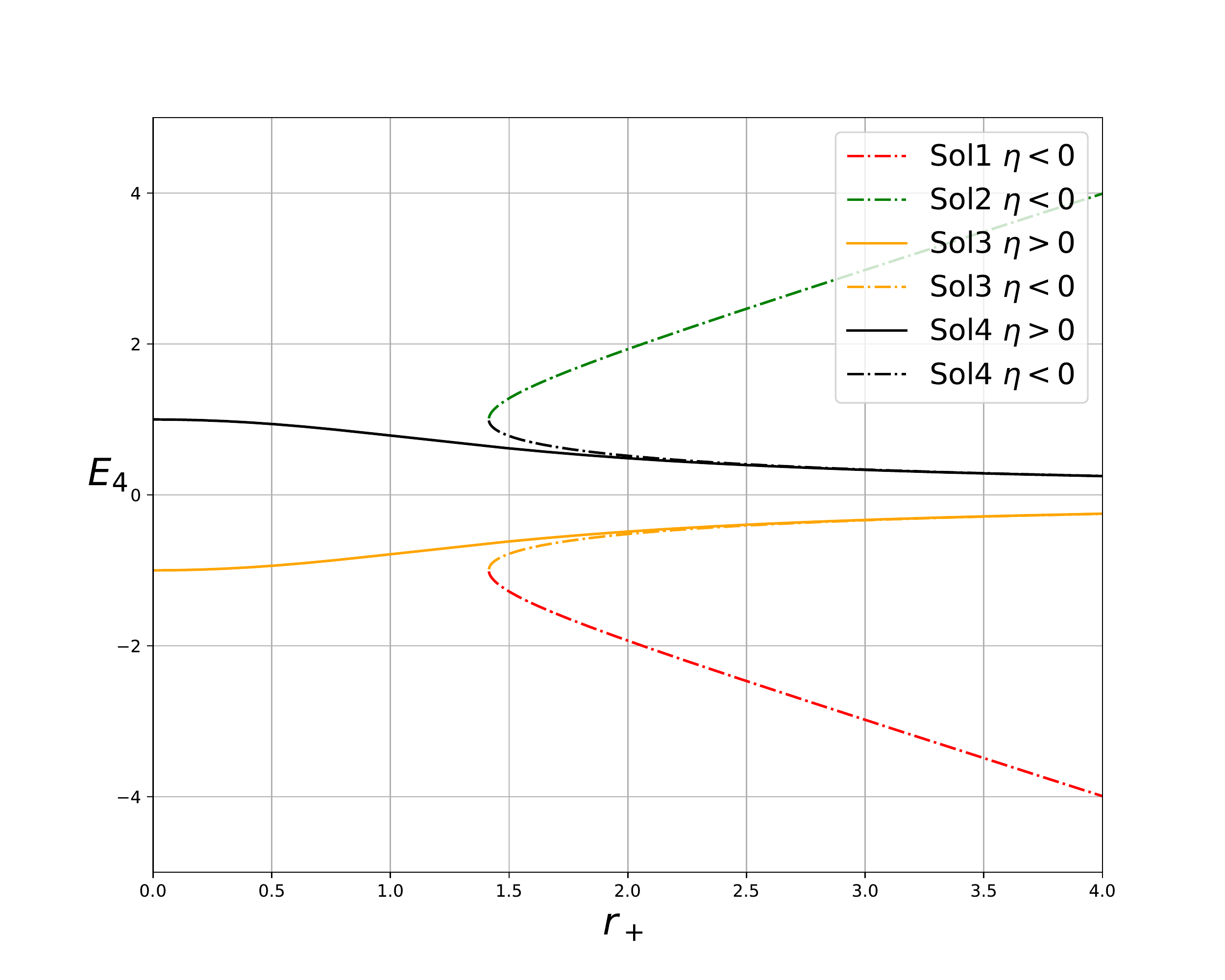}
         \caption{$n=4, |\eta| = 0.9$}
     \end{subfigure}
     \caption{Plot of energy by $r_+$ for all the solutions of the modified dispersion relation, for the cases $n=1$ and $n=4$.}
     \label{fig:MDR}
\end{figure}

The temperature of the black hole, given by the equation (\ref{Temp}), will depend not only on the parameters of the metric but also on the energy present in the Rainbow functions. This is the energy of the particles that "feels" the metric; in this case, the energy of the virtual particles emitted by the black hole. For the sake of simplicity, we will consider that only photons are emitted.

Since the Hawking temperature must contain only parameters of spacetime, we have to relate the energy of the photons to the parameters of the black hole that radiates them. We will do this considering that the photons form an ensemble with a well-defined temperature. The energy of the photons is given by the MDR, equation (\ref{mdr}), and we will relate the momentum of a photon with the horizon radius through the Heisenberg uncertainty relation, $p = 1/r_+$, using $\hbar = 1$. 

We would like to draw attention to the fact that the author in 
\cite{Ali:2014xqa} used $p = E$ in the Heisenberg uncertainty relation. Using this dispersion relation is the reason why the authors found black holes remnants in their studies. This issue was mentioned and corrected in \cite{Gim:2014ira}, where the correct MDR was used, but many authors still claim that Rainbow gravity allows remnants based on the results found in \cite{Ali:2014xqa}. As we will see, our study found no evidence of remnants in Rainbow.

Note that the MDR, equation (\ref{mdr}), is a polynomial in the energy. The usual dispersion relation is a quadratic polynomial and gives us two values for the energy, one positive and another negative. For the MDR, and especially for $n > 2$, there will be more than two solutions, and we need to verify which of them are physically acceptable.

For $n=0$, there is only one positive solution, as long as $\eta < 1$. For $n=1$, one can find an analytical solution,

\begin{equation}
    E_{n=1} = \frac {-\eta+\sqrt {{\eta}^{2}+4\,{r_+}^{2}}}{2 \, r_+^2}
\end{equation}

\noindent 
where $\eta$ can be positive or negative. For $n=2$, we have

\begin{equation}
    E_{n=2} = \pm \frac {1}{\sqrt {{r_+}^{2}+\eta}}
\end{equation}

\noindent
so that $\eta > - r_+^2$, and for $n=4$, there are four solutions, and their expressions are too wide for us to show here; instead, we will depict it. In figure (\ref{fig:MDR}), we plot both $n=1$ and $n=4$ solutions. 

For $n=1$ we have two positive solutions; the first one goes to infinity as $r_+ \rightarrow 0$, the same behavior of the usual Schwarzschild black hole. The second one introduces a new behavior and, as $r_+ \rightarrow 0$, the energy approaches a finite value. The same also happens for $n=2$.

For $n=4$ we have three real positive solutions, but only one is valid for all $r_+$. Also, there is an interchange of solutions for $\eta < 0$ that should be better analyzed when we study black hole thermodynamics. Note that this solution appears to be continuous, so one can consider it as a valid real solution described by a piecewise function, even though it does not cover all values for the horizon radius. 

\subsection{Horizons in a Kiselev Black Hole}

Before proceeding to study thermodynamics in Rainbow Gravity, we will analyze the horizons in a Kiselev Black Hole. Such horizons will be the same both in GR as in Rainbow, since it does not depend on the Rainbow functions. The horizons are given by $g_{tt} = 0$, that results in the equation (\ref{eqMass}). As one can see, positive values for the mass will depend on a relation between the parameters of the theory and the horizon radius, $r_+$.

\begin{figure}[h]
         \centering
         \includegraphics[width=0.70
         \textwidth]{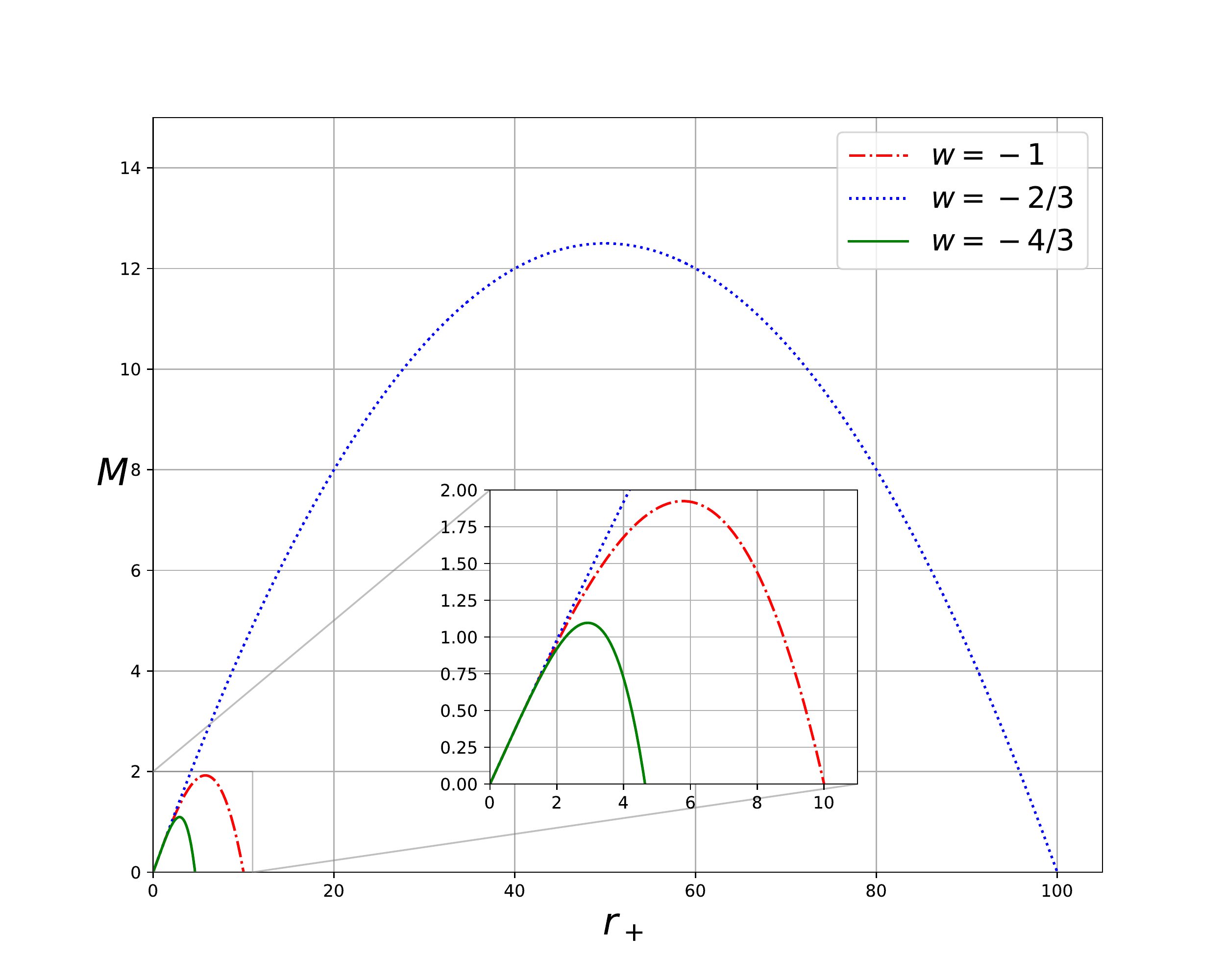}
     \caption{Plot of mass by $r_+$ for $\omega = -1, -2/3$ and $-4/3$. In the center of the graph, a closer depict for small values of the horizon radius. We used $c = 0.01, G = 1/E_p^2 = 1$ for the rest of the discussion.}
     \label{fig:Mass}
\end{figure}

In figure (\ref{fig:Mass}), we depict this relation for some values of $\omega$, and we will set $G = 1/E_p^2 = 1$ and $c=0.01$ for the rest of the discussion. As one can see, the mass has an upper bound that corresponds to the case where two horizons coincide: the black hole horizon and the cosmic horizon. This kind of black hole is usually called Nariai black hole when the cosmic fluid is a cosmological constant. 

As the mass decreases, the horizons start to move away from each other; the left horizon is the black hole horizon while the right one is the cosmological horizon. This means that the maximum value for the black hole horizon is the point where the mass reaches its critical point; above this value either we are after the cosmological horizon or we have a naked singularity.

We are calling attention to this point because it will be relevant for the study of the thermodynamics of the system. As we will see, this is exactly the point where the temperature will reach a critical value and cannot be extended for larges values of $r_+$. 
\subsection{Temperature, Entropy and Specific Heat}

The temperature of the black hole in Rainbow gravity is given by

\begin{equation}
    T_H = \frac{r_+^{-3 \omega - 2} \, |3\,c\, \omega + r_+^{3\omega + 1}|}{4 \pi} \left(1 - \eta \left(\frac{E}{E_p}\right)^n \right)^{1/2} 
\end{equation}

\noindent
and we need to find how $E$ relates to $r_+$. For this, we will use the uncertainty relation and the dispersion relation we already found. 

As we have shown, for $n=0$, $n=1$ and $n=2$, we have only one positive solution for the energy, for each value of the parameter $\eta$ (positive or negative). The case $n=4$ is more subtle, since we have two solutions for $\eta < 0$, but that apparently form a piecewise continuous function.

Entropy can be integrated as

\begin{equation}
    \label{entropy}
    S =\int \frac{1}{T} \frac{d M}{d T} dT = \int \frac{1}{T} \frac{dM}{dr_+} dr_+
\end{equation}

\noindent
and the specific heat can be derived as

\begin{equation}
    C \coloneqq \frac{dM}{dT} = \frac{dM}{dr_+} \left( \frac{dT}{dr_+} \right)^{-1}.
\end{equation}

One can find an analytic solution for the specif heat, but the formula will be lengthy and will depend on the choice of the parameter $n$. Instead of s lengthy formula, in the next section we will depict several graphs for the entropy and specific heats. 

\section{Temperature, entropy, specific heat and remnants}

Let us now proceed to study the thermodynamics of black holes, in Kiselev spacetime, in the framework of Rainbow gravity.

\subsection{Temperature}

\begin{figure}
     \centering
     \begin{subfigure}[b]{0.45\textwidth}
         \centering
         \includegraphics[width=1.2\textwidth]{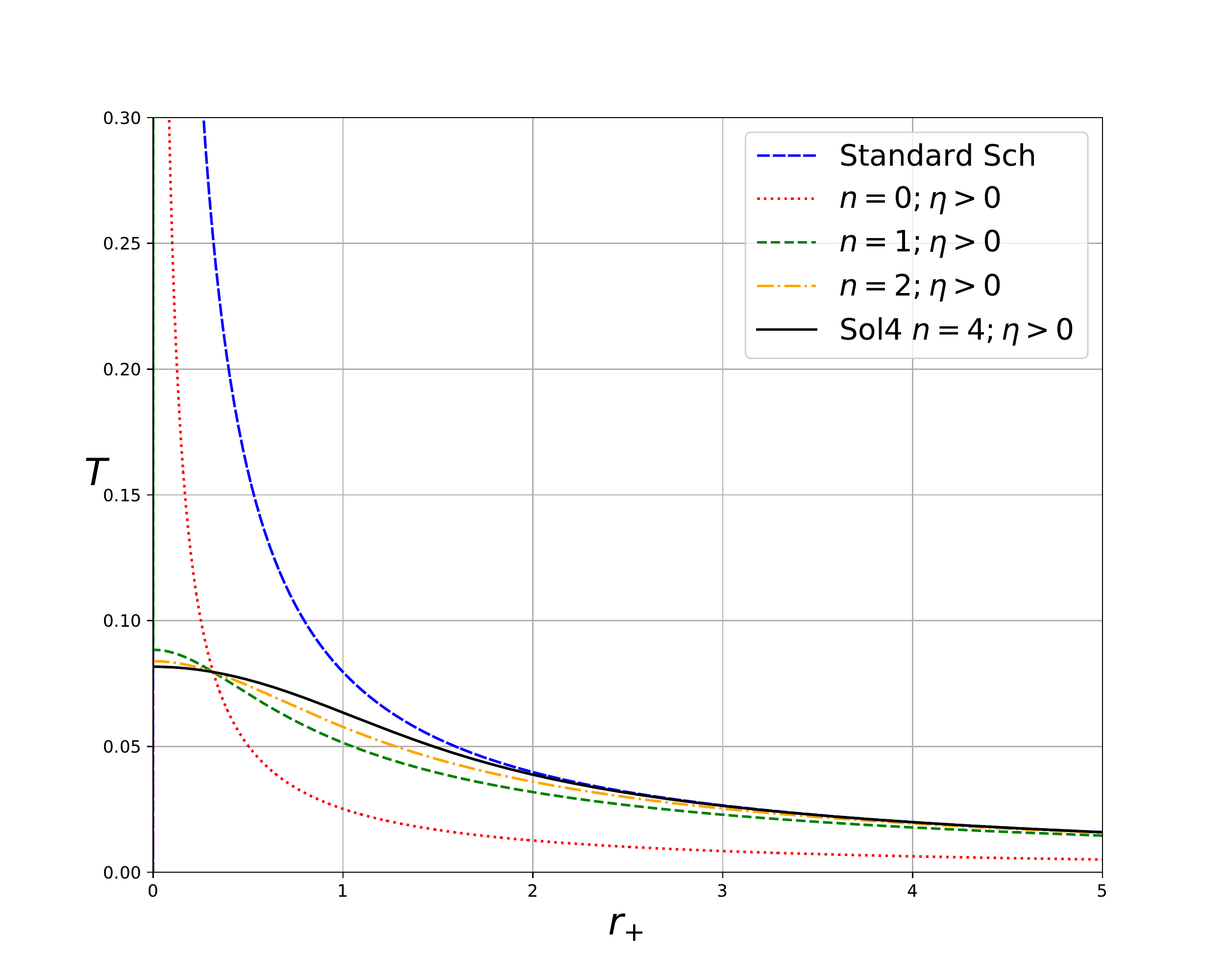}
         \caption{$\eta = 0.9$}
     \end{subfigure}
     \hfill
     \begin{subfigure}[b]{0.45\textwidth}
         \centering
         \includegraphics[width=1.2\textwidth]{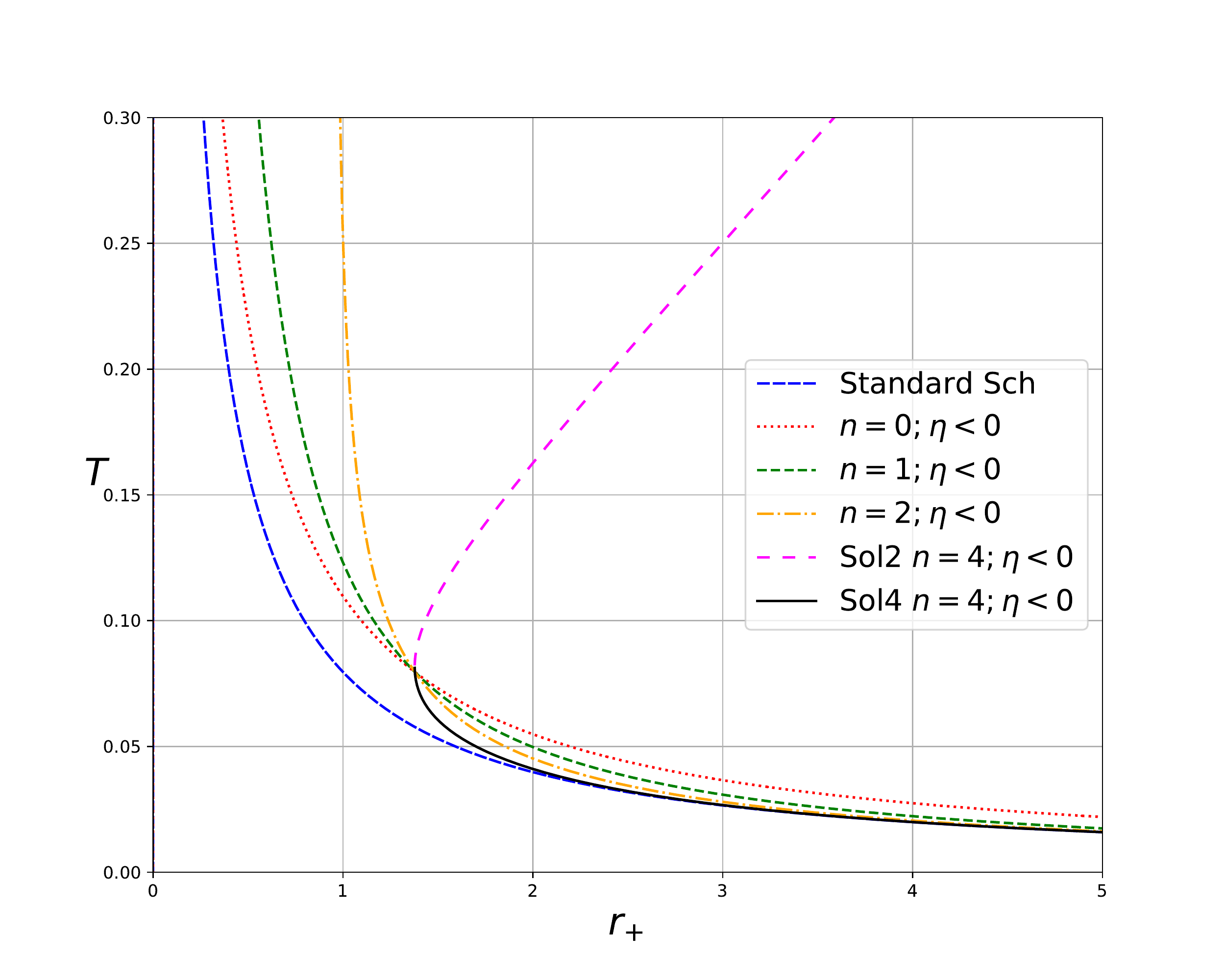}
         \caption{$\eta = -0.9$}
     \end{subfigure}
     \hfill
        \caption{Temperature as a function of the horizon for Rainbow's gravity without a cosmic fluid.}
        \label{fig:Temperature}
\end{figure}

Let's start by ignoring the cosmic fluid and doing a preliminary study of black holes thermodynamics in Rainbow gravity. This is not a new topic, but we hope to clarify some issues that are not clear in the literature. For that, we will look for solutions for different values of the exponent $n$ and explicitly depicted the temperature as a function of the horizon radius, as show in figure (\ref{fig:Temperature}). We also plot the temperature for the Schwarzschild black hole to compare Rainbow gravity with General Relativity.

For $\eta > 0$, the temperature blows up for General Relativity and also for $n=0$, otherwise it approaches a finite value. The shape of the curve is different for different exponents, but their behavior is similar.

For $\eta<0$, all studied cases until $n<4$ leads to an infinite temperature as the radius approaches zero, which is consistent with the energy of the photons (or any massless particle) near the black hole when one uses these MDRs. For $n=4$ we have two solutions with a bound to the horizon radius. At this bound, the temperature is not zero, so it should continue radiating in a kind of bouncing behavior. This is an interesting new behavior, since the radius can first decrease and then increase (or the reverse), but the temperature will always decrease (or always increase).   

\begin{figure}[!h]
         \centering
         \includegraphics[width=0.70
         \textwidth]{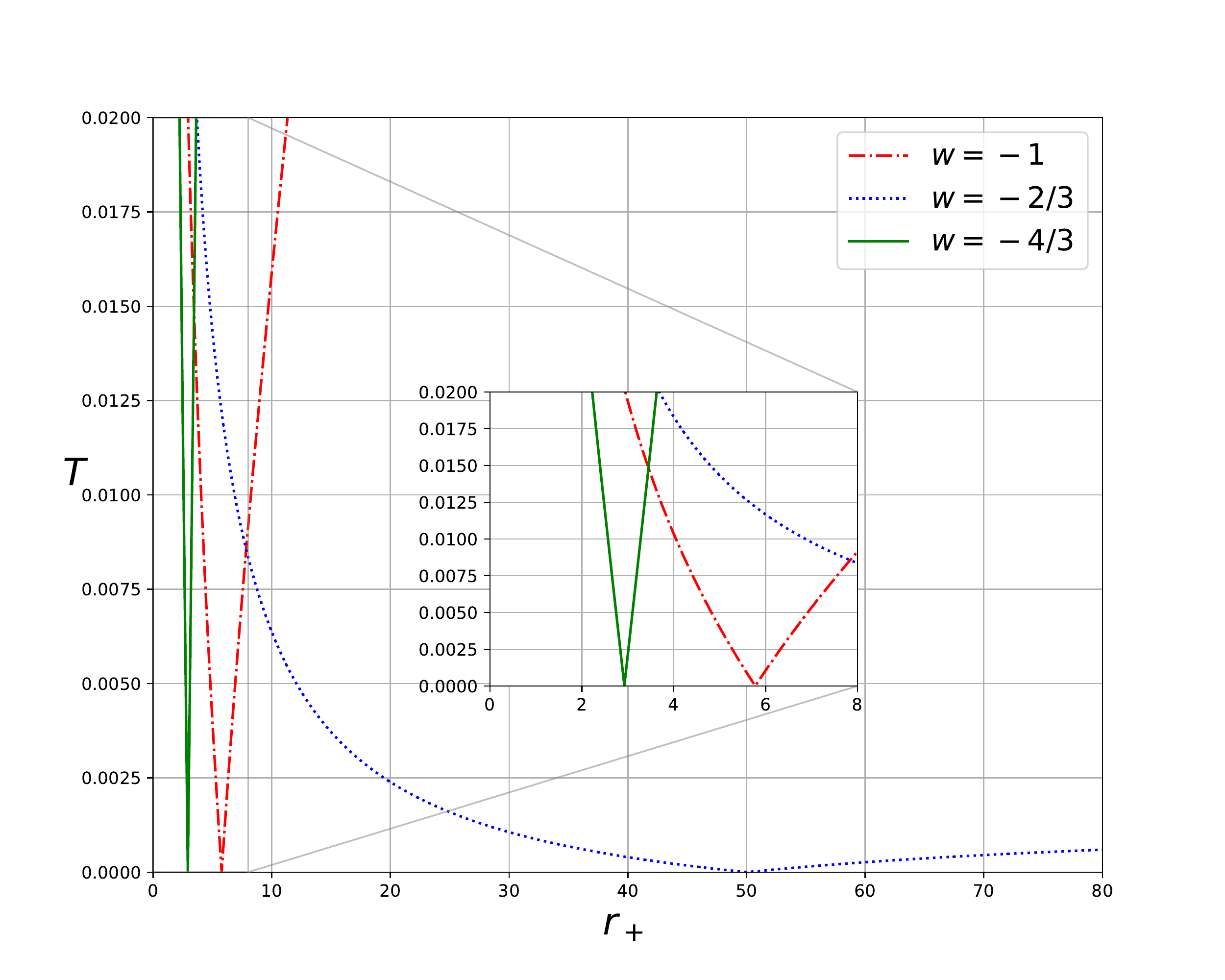}
     \caption{Temperature as a function of the horizon radius for Kiselev black holes, in general relativity.}
     \label{fig:TempKiselev}
\end{figure}

We will now briefly study the temperature of the black hole in Kiselev's spacetime, disregarding the Rainbow functions. The idea is to identify this behavior from the one due to Rainbow. The temperature is depicted in figure (\ref{fig:TempKiselev}), and one can see that there is a point where the temperature is zero. This point is precisely that of critical mass, where the horizons of the black hole and cosmological coincide.

This occurs for all values of $\omega$ studied, and what varies is only the critical horizon radius. As mentioned before, for horizon radius values greater than this critical value, we are either considering the cosmological horizon or what we have is a naked singularity. In any case, it is not an acceptable physical situation.

\begin{figure}[!h]
     \centering
     \begin{subfigure}[b]{0.45\textwidth}
         \centering
         \includegraphics[width=1.2\textwidth]{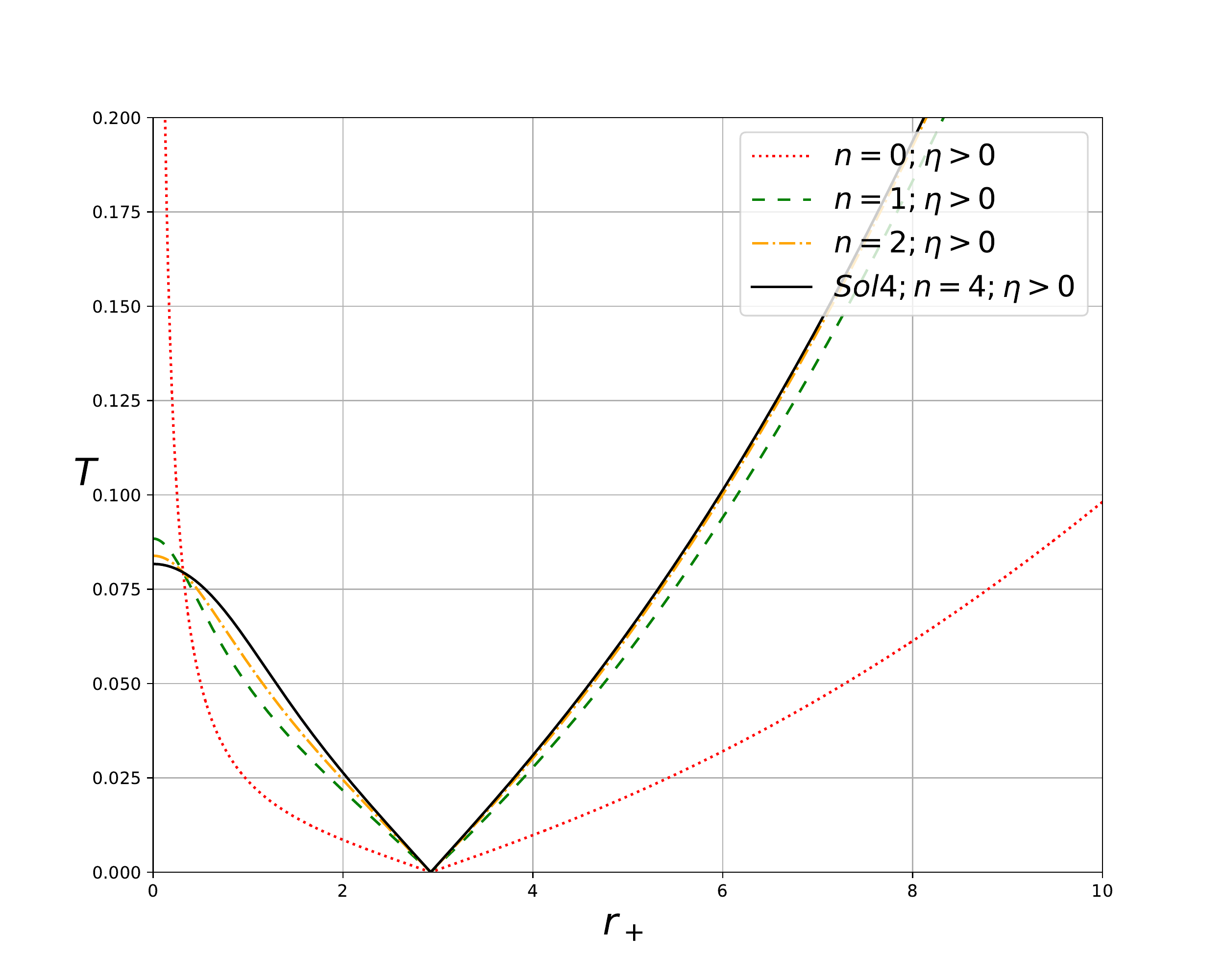}
         \caption{$\eta = 0.9$}
     \end{subfigure}
     \hfill
     \begin{subfigure}[b]{0.45\textwidth}
         \centering
         \includegraphics[width=1.2\textwidth]{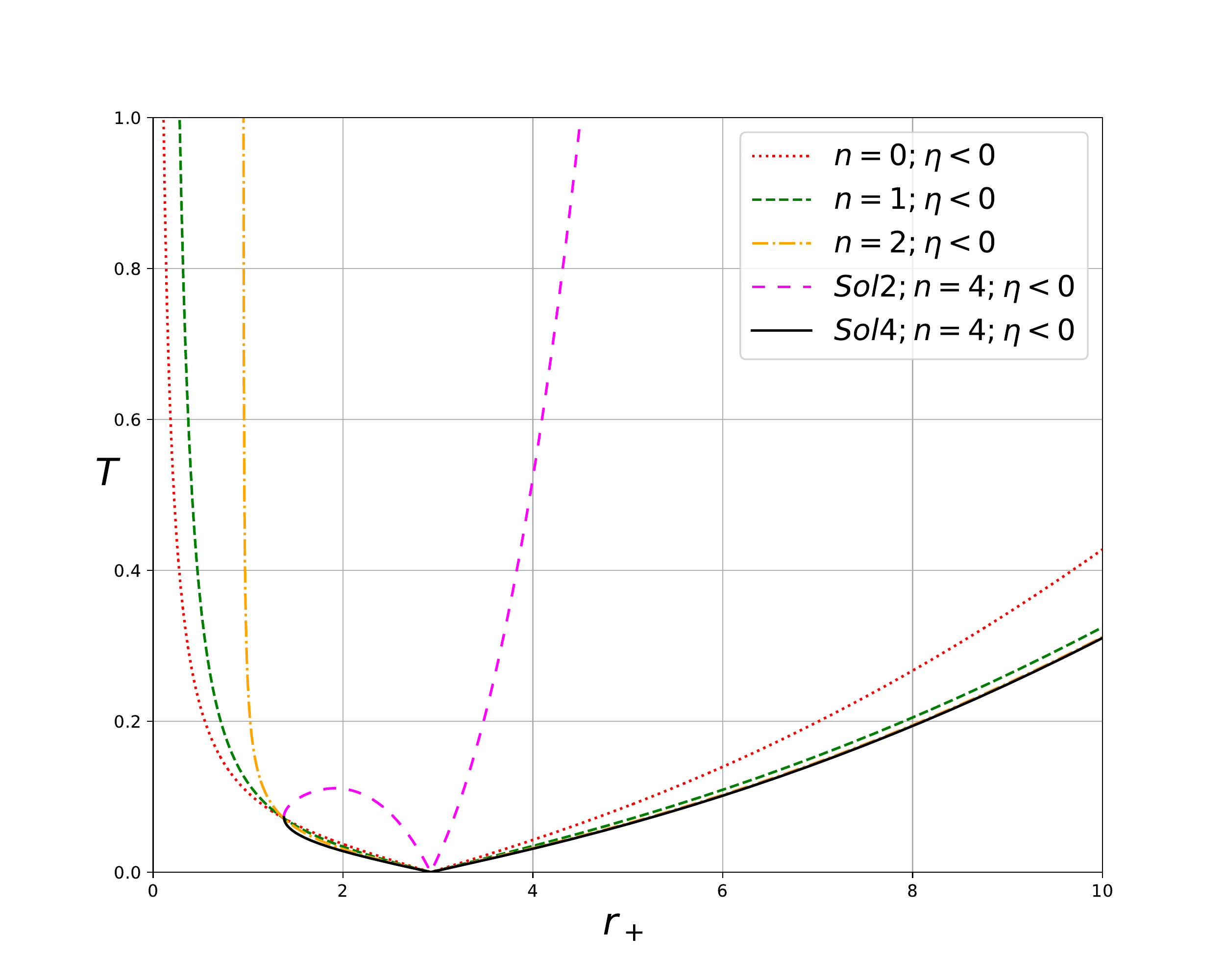}
         \caption{$\eta = -0.9$}
     \end{subfigure}
     \hfill
        \caption{Temperature as function of the horizon radius, for Kiselev black holes, $\omega = -4/3 $, in Rainbow gravity.}
        \label{fig:TempKiselevRainbow}
\end{figure}

Let us now consider the complete case of Kiselev black holes in Rainbow gravity. Figure (\ref{fig:TempKiselevRainbow}) shows the temperature of the black hole, for $\omega = - 4/3$, for different values of exponent $n$. On the left, we have the case $\eta < 0$, and on the right, the case $\eta > 0$. As we can see, what happens is a union of both behaviors; due to the incorporation of a cosmic fluid, the temperature will now go to zero in a critical horizon radius, instead of going to infinity, and the black hole horizon must be smaller than such a critical horizon. For $n < 4$, as the radius of the horizon decreases, the temperature can go to infinity or a limit value. Either way, it doesn't go to zero. This is important for the study of remnants.

The exotic case occurs when $n = 4$, when the temperature presents a kind of loop, which indicates that the temperature can only decrease from a finite value to zero, thus leaving a remnant. However, the final value of the horizon radius would be equal to the cosmological horizon, which represents an extreme situation; whether such a scenario is physically feasible or not is a difficult question to answer.

\begin{figure}[!h]
     \centering
     \begin{subfigure}[b]{0.45\textwidth}
         \centering
         \includegraphics[width=1.2\textwidth]{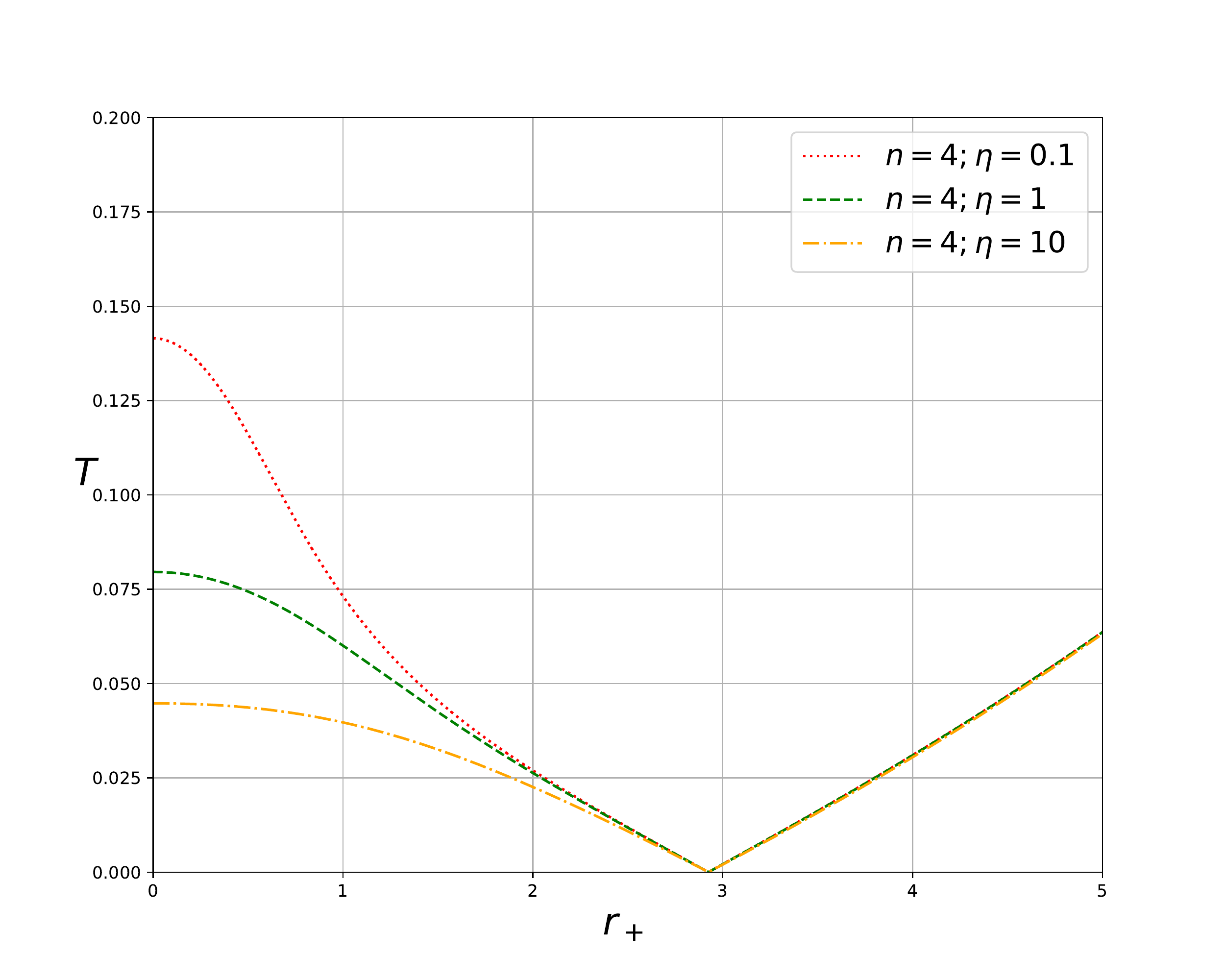}
         \caption{$\eta > 0$}
     \end{subfigure}
     \hfill
     \begin{subfigure}[b]{0.45\textwidth}
         \centering
         \includegraphics[width=1.2\textwidth]{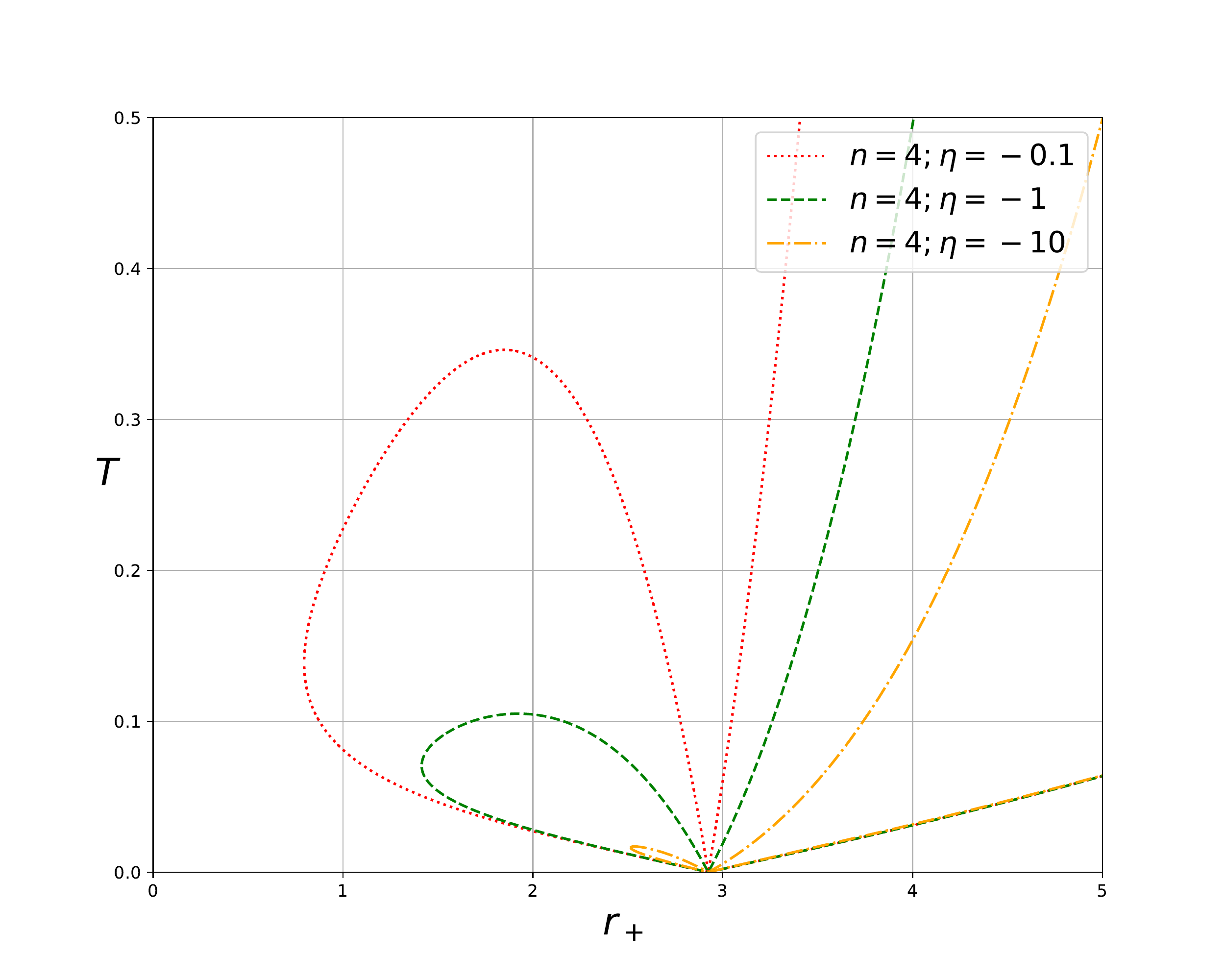}
         \caption{$\eta < 0$}
     \end{subfigure}
     \hfill
        \caption{Temperature as a function of the horizon radius, for Kiselev black holes, $\omega=-4/3$, in Rainbow gravity, for a fixed exponent $n=4$. In this graph, we vary the parameter $\eta$.}
        \label{fig:TempKiselevRainbowEta}
\end{figure}

In figure (\ref{fig:TempKiselevRainbowEta}), we depict the behavior of the temperature for different values of the parameter $\eta$, for a fixed exponent $n = 4$. As one can see, the behavior is similar, but the loop size for $\eta < 0$ and the final temperature value for $\eta > 0$ are different. For other values of the exponent, we found a similar result, always maintaining the same behavior, but varying the path in the $T-r_ +$ plane.

\begin{figure}[!h]
     \centering
     \begin{subfigure}[b]{0.45\textwidth}
         \centering
         \includegraphics[width=1.2\textwidth]{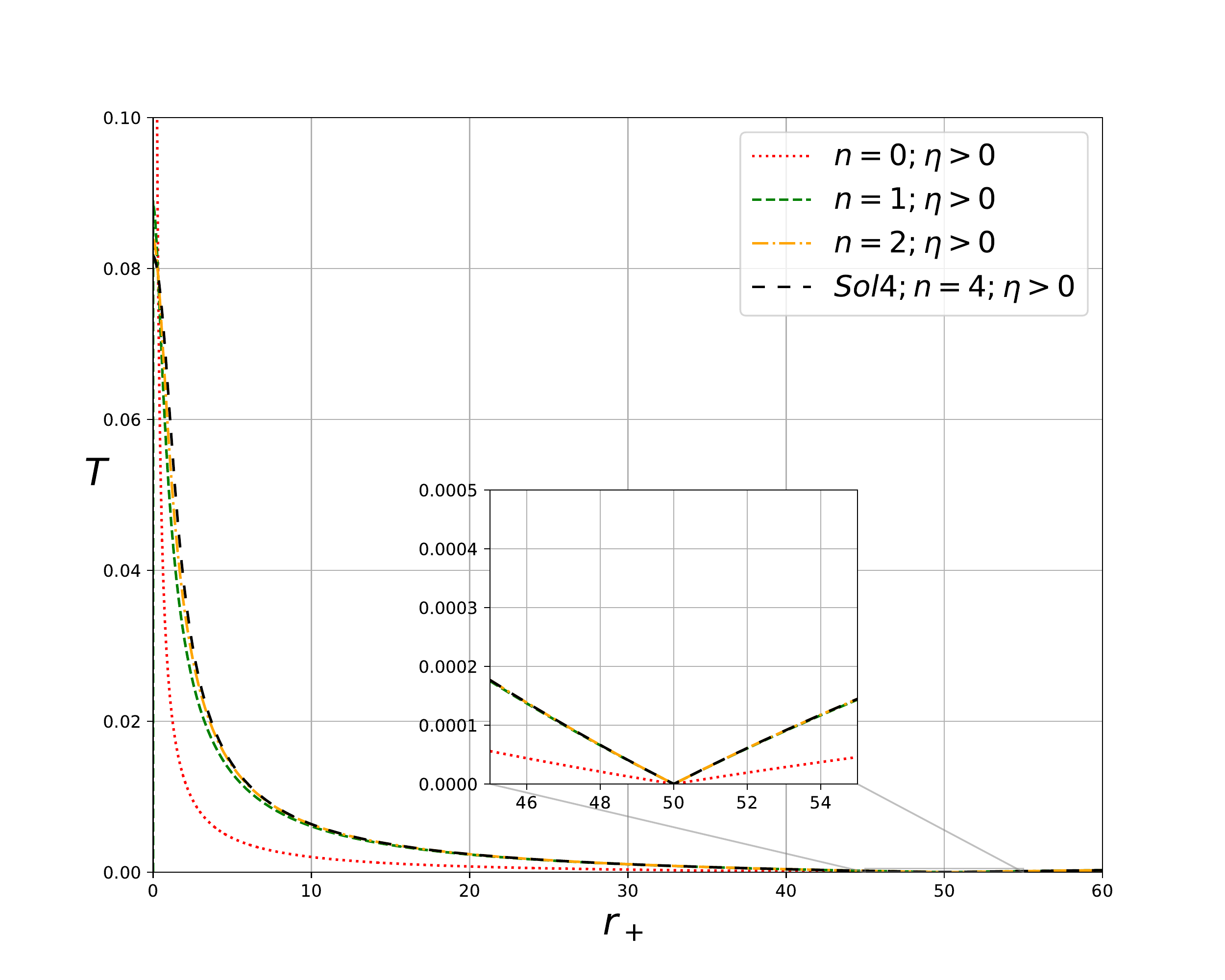}
         \caption{$\eta = 0.9$}
     \end{subfigure}
     \hfill
     \begin{subfigure}[b]{0.45\textwidth}
         \centering
         \includegraphics[width=1.2\textwidth]{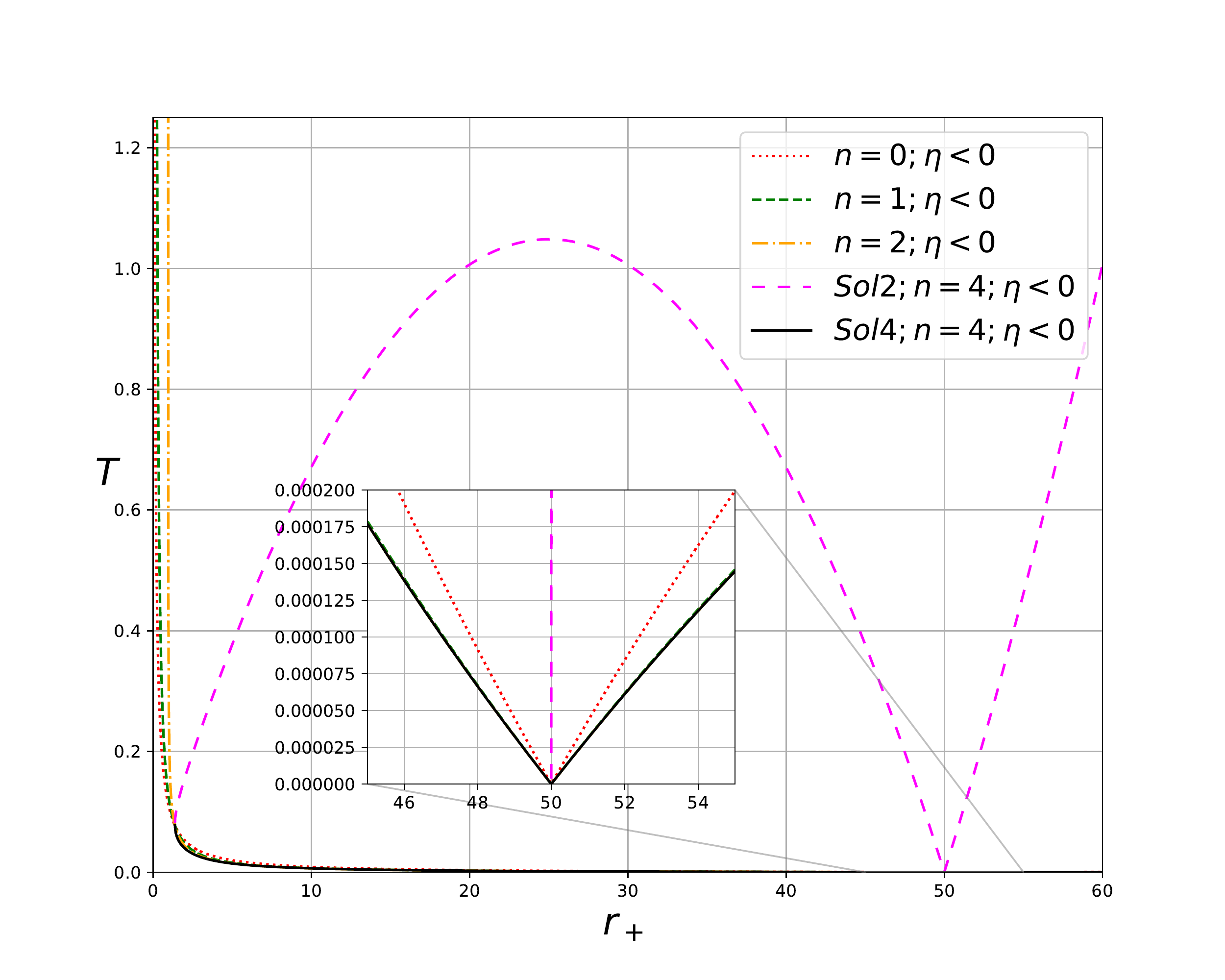}
         \caption{$\eta = -0.9$}
     \end{subfigure}
     \hfill
        \caption{Temperature as a function of the horizon radius, for Kiselev black holes, $\omega=-2/3$, in Rainbow gravity.}
        \label{fig:TempKiselevRainbow32}
\end{figure}

Finally, in figure (\ref{fig:TempKiselevRainbow32}), we plot the temperature for $\omega = -2/3$. The result is also similar to the case $\omega = -4/3$, but because the critical horizon radius is larger, the plot is less clear than the case of phantom energy. This was the reason we choose to use $\omega = -4/3$ in most cases. For a cosmological constant the behavior is also similar and that is the reason we will not show it here.

\subsection{Entropy}

The entropy of a black hole, in the case of general relativity, is given by Bekenstein's formula, $S = \pi r_+^2$ \cite{Bekenstein:1973ur}. The horizon radius clearly will depend on the parameter of the spacetime, but as long we are interested on the values of the entropy as a function of the horizon radius, the formula will not receive any correction. This means that, for Kiselev black holes, the only effect will be to limit the value for the horizon radius, as one can see in figure (\ref{fig:EntropyKiselev}).

\begin{figure}[!h]
         \centering
         \includegraphics[width=0.70
         \textwidth]{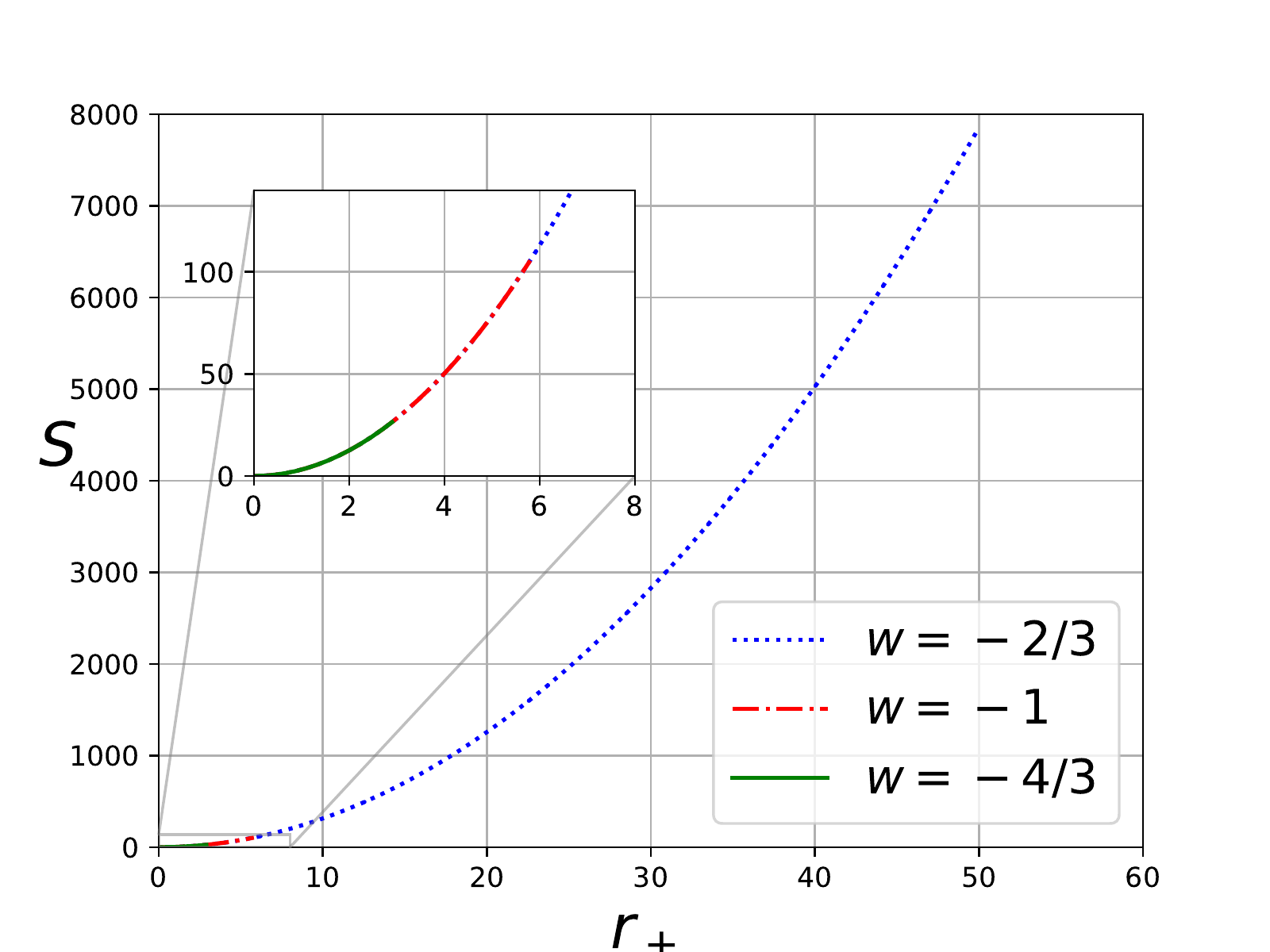}
     \caption{Entropy as function of the horizon radius, for Kiselev black holes, in general relativity.}
     \label{fig:EntropyKiselev}
\end{figure}

The same will not hold in Rainbow, since we cannot consider that the Bekenstein's formula is still valid. To plot the graphs, we will use the first law of thermodynamics that gives us the entropy as equation (\ref{entropy}). 

\begin{figure}[!h]
     \centering
     \begin{subfigure}[b]{0.45\textwidth}
         \centering
         \includegraphics[width=1.2\textwidth]{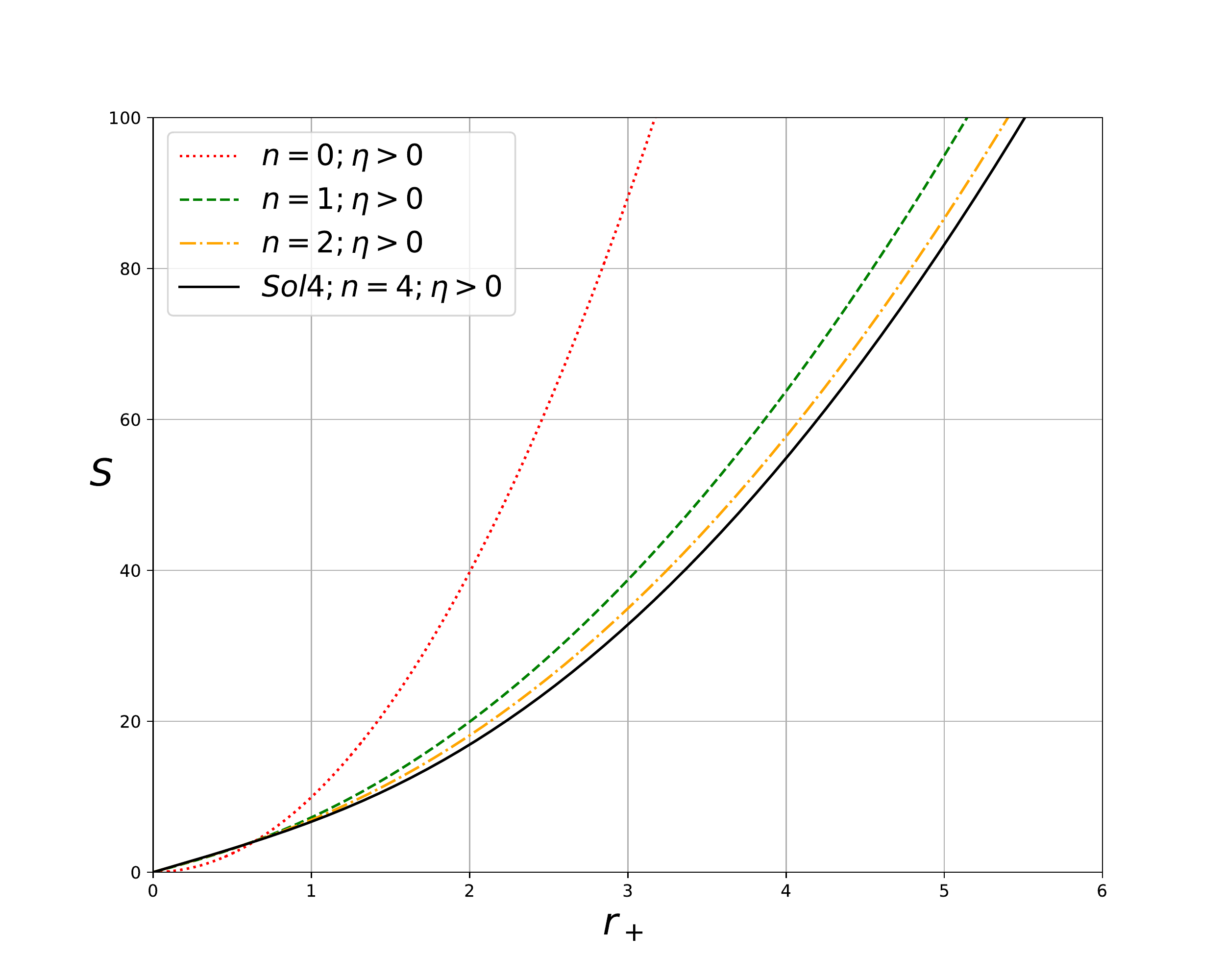}
         \caption{$\eta = 0.9$}
     \end{subfigure}
     \hfill
     \begin{subfigure}[b]{0.45\textwidth}
         \centering
         \includegraphics[width=1.2\textwidth]{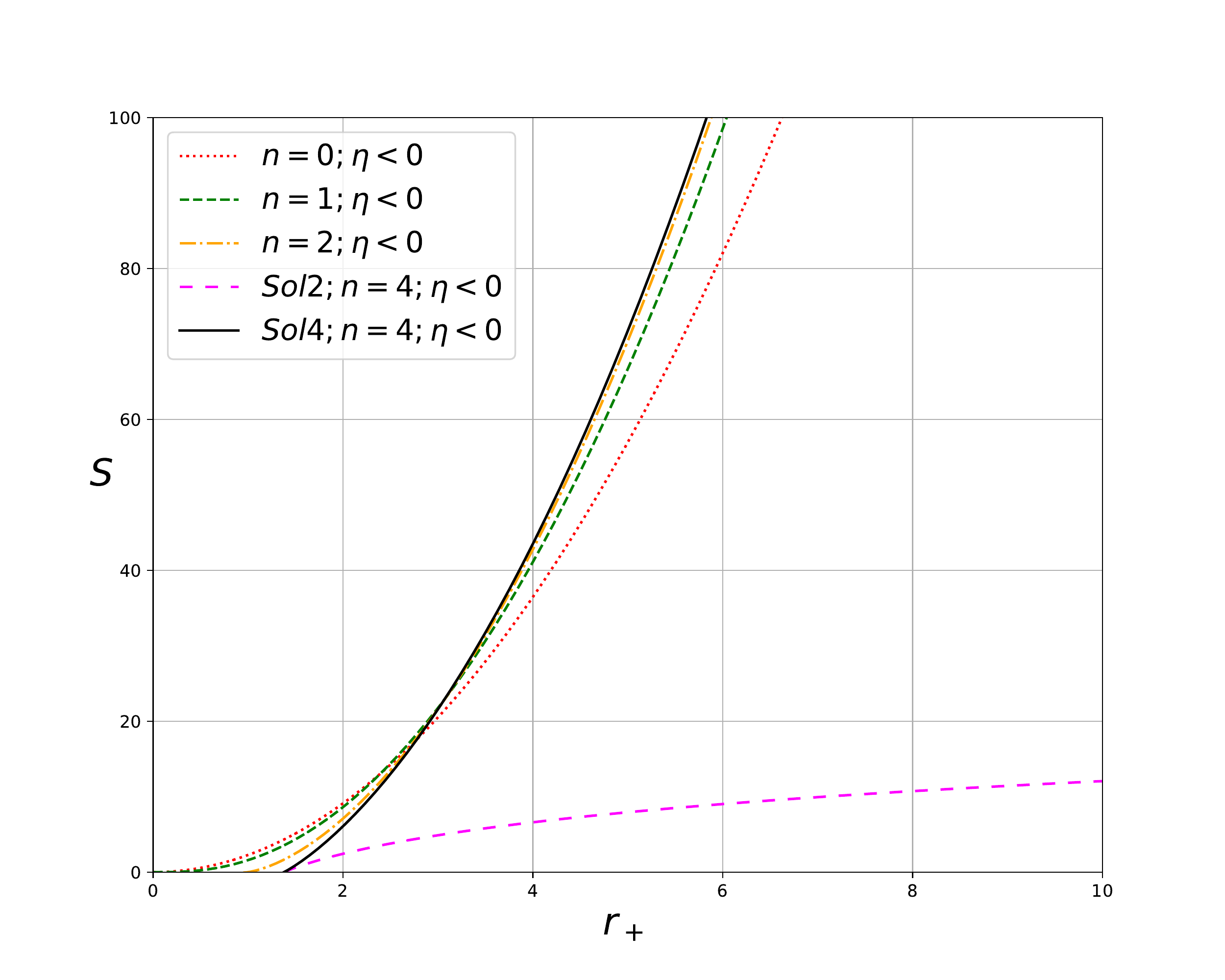}
         \caption{$\eta = -0.9$}
     \end{subfigure}
     \hfill
        \caption{Entropy as a function of the horizon radius in Rainbow gravity.}
        \label{fig:entropyKiselev}
\end{figure}

That Rainbow gravity leads to a correction in entropy had already been proposed in \cite{Ling:2005bp}, and such corrections are explicitly depicted in figure (\ref{fig:entropyKiselev}), without considering the cosmic fluid. For some values of the exponent $n$, we can found an analytic formula for the entropy, but not for all values of the exponent. For $n=4$, for example, the entropy need to be calculated numerically. 

When we consider the complete case of Kiselev black hole in Rainbow gravity, these behaviors are complementary; the shape of the graph does not change, but the horizon radius is limited. Because of this, there is no need to plot the graphs.

\subsection{Specific Heat}

We are now going to study the specific heat. For the Schwarzschild black hole, in general relativity, the specific heat is negative for all values of the horizon radius. The same is a reflection of the fact that the temperature decreased with increasing radius. 

\begin{figure}[!h]
         \centering
         \includegraphics[width=0.70
         \textwidth]{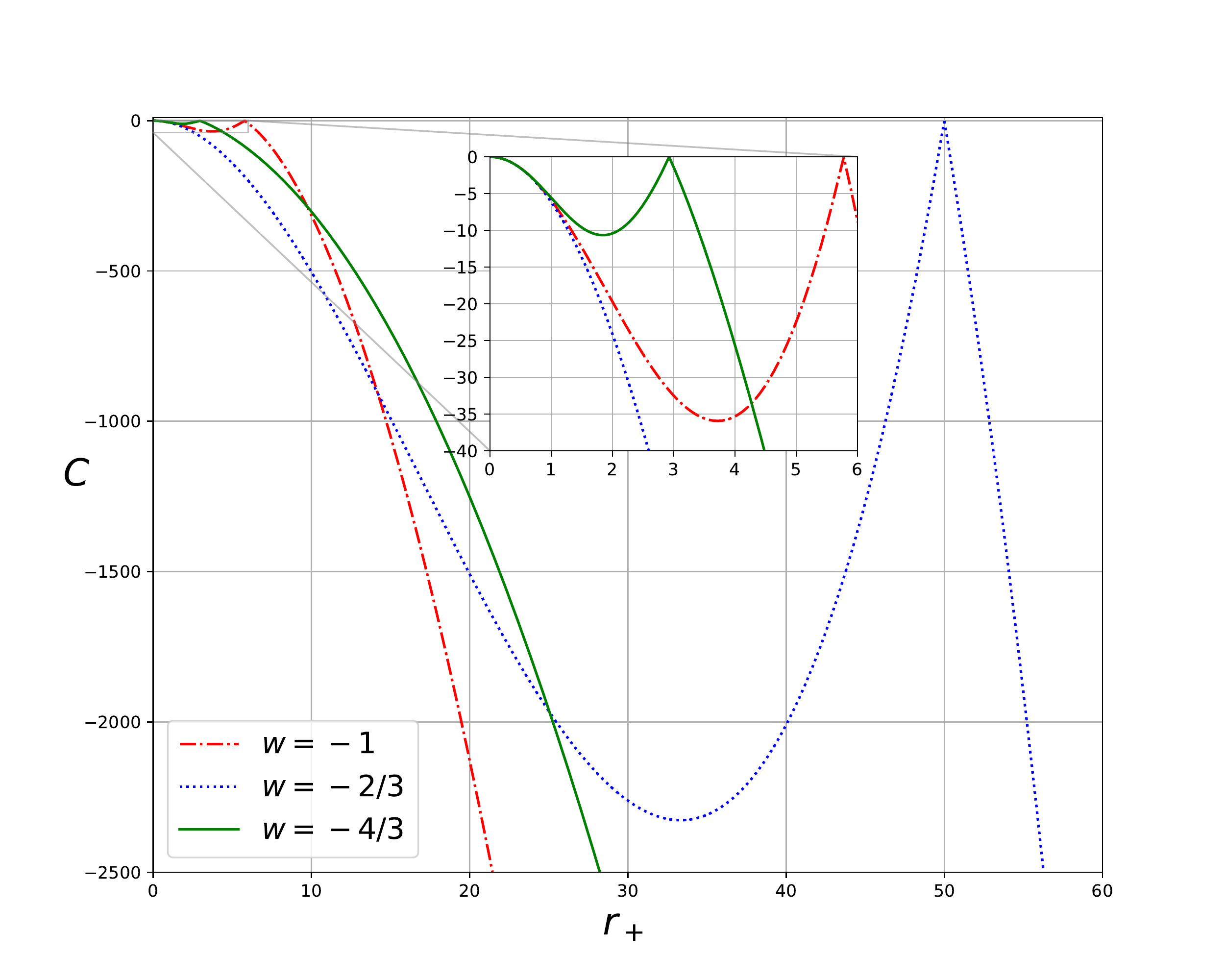}
     \caption{Specific heat as function of the horizon radius, for Kiselev black holes, in general relativity.}
     \label{fig:EntropyKiselev}
\end{figure}

For Kiselev black hole the same will hold, but with a different behavior. In figure (\ref{fig:EntropyKiselev}), we depict it for the cases of the parameter $\omega = -1, -2/3$ and $-4/3$. The specific heat is still negative, but now it has an inflection point, since the horizon radius has a limit.  

\begin{figure}[!h]
     \centering
     \begin{subfigure}[b]{0.45\textwidth}
         \centering
         \includegraphics[width=1.2\textwidth]{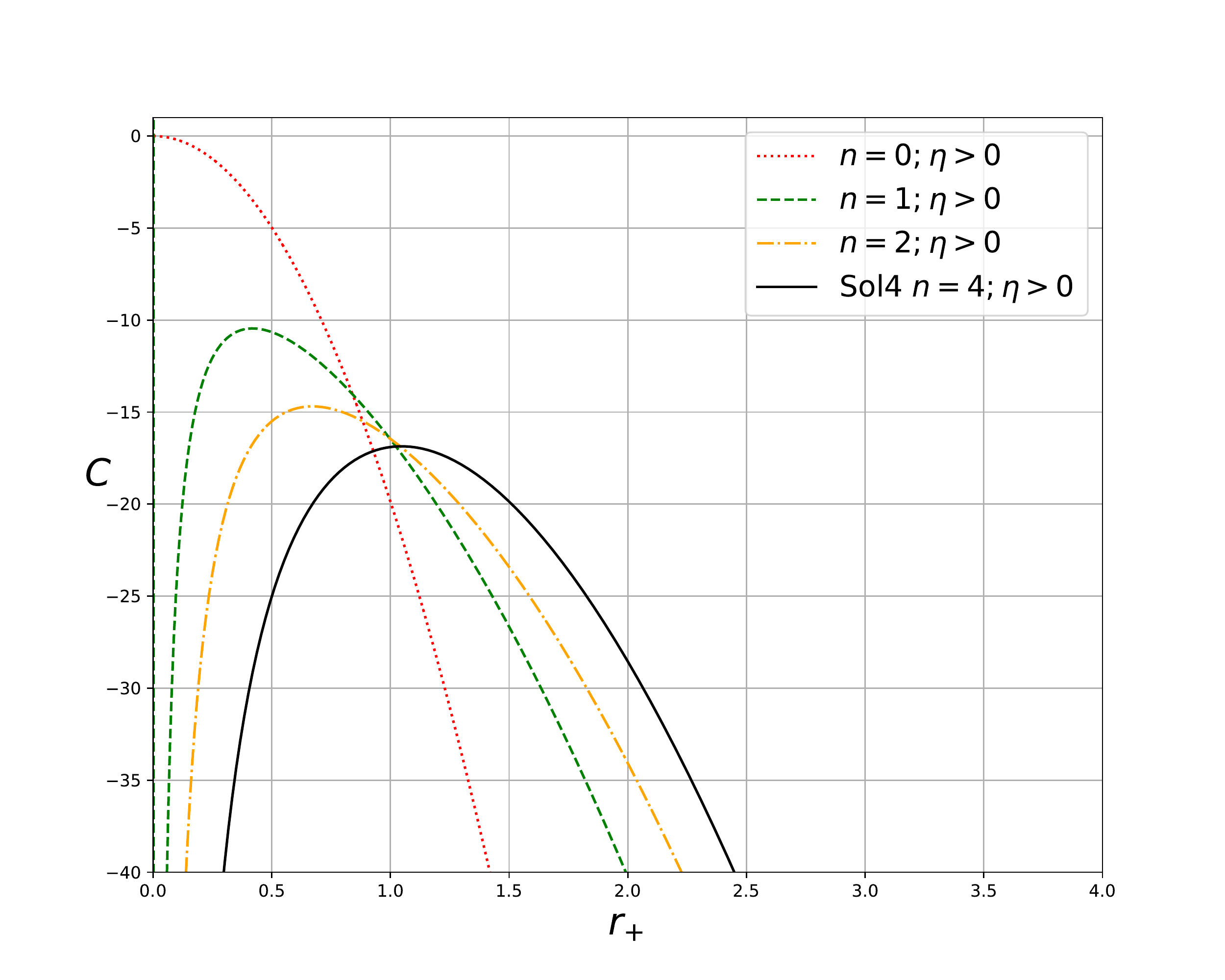}
         \caption{$\eta = 0.9$}
     \end{subfigure}
     \hfill
     \begin{subfigure}[b]{0.45\textwidth}
         \centering
         \includegraphics[width=1.2\textwidth]{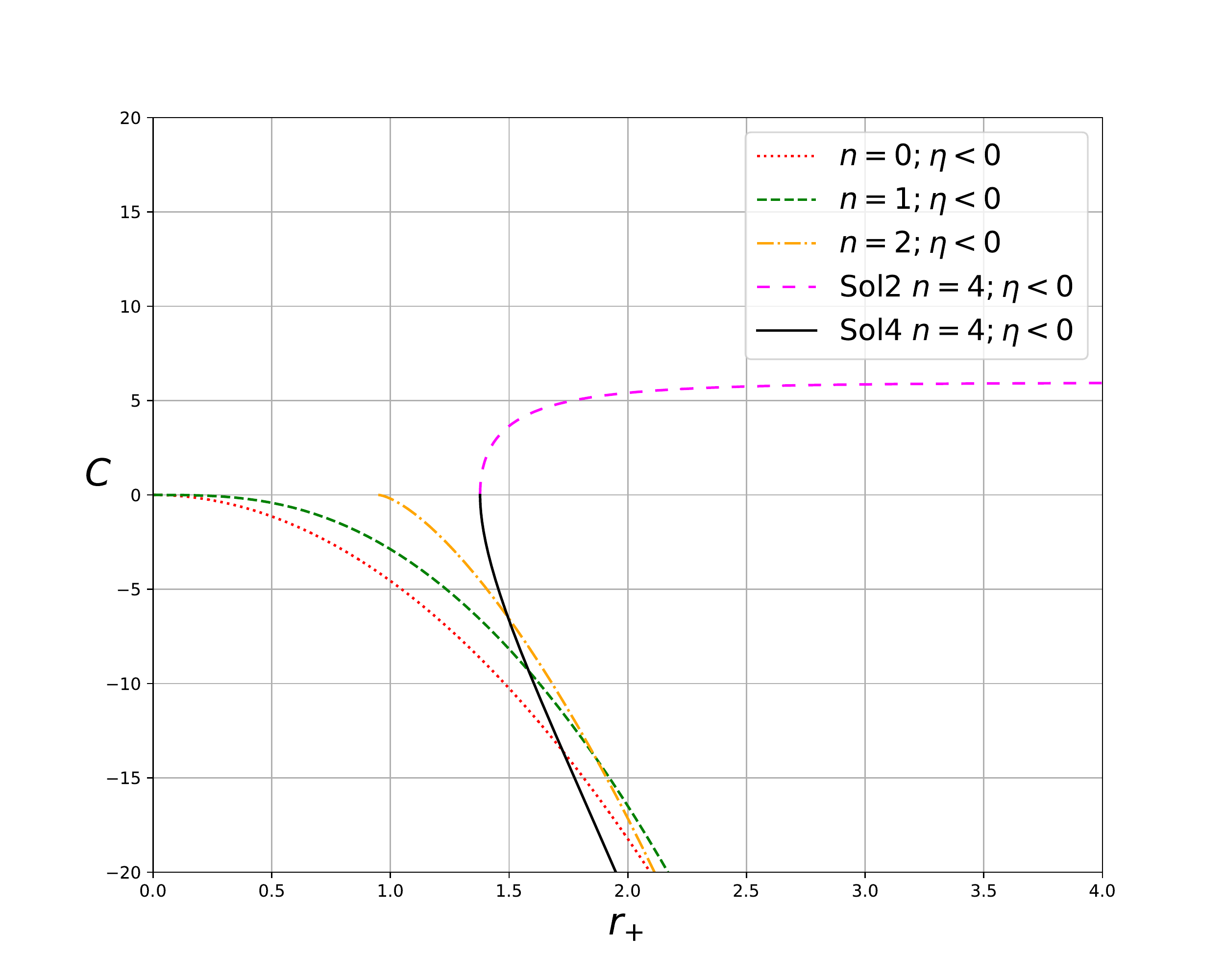}
         \caption{$\eta = -0.9$}
     \end{subfigure}
     \hfill
        \caption{Specific heat as function of the horizon radius, without any cosmic fluid, in Rainbow gravity.}
        \label{fig:specificHeatRainbow}
\end{figure}

We are now going to study the specific heat in Rainbow, without the cosmic fluid. Note that when the temperature tends to infinity, the specific heat tends to zero. When the temperature tends towards a finite value, the specific heat tends towards minus infinite. This is depicted in figure (\ref{fig:specificHeatRainbow}). 

The possible explanation should follow the following lines. A very small specific heat allows a large increase in temperature, even for a small variation in the heat exchanged with the environment. Conversely, a very large specific heat (in absolute value) will result in a small variation in temperature. Thus, the temperature and specific heat graphs are consistent.

A new case occurs for $ n=4 $, when the model has positive specific heat, a characteristic not very common for black holes. This is because, in a certain region, the temperature of the black hole can decreases with the horizon radius. After reaching a critical value, the behavior changes, and the temperature starts to decrease as the horizon radius increases (or the other way around).

\begin{figure}[!h]
     \centering
     \begin{subfigure}[b]{0.45\textwidth}
         \centering
         \includegraphics[width=1.2\textwidth]{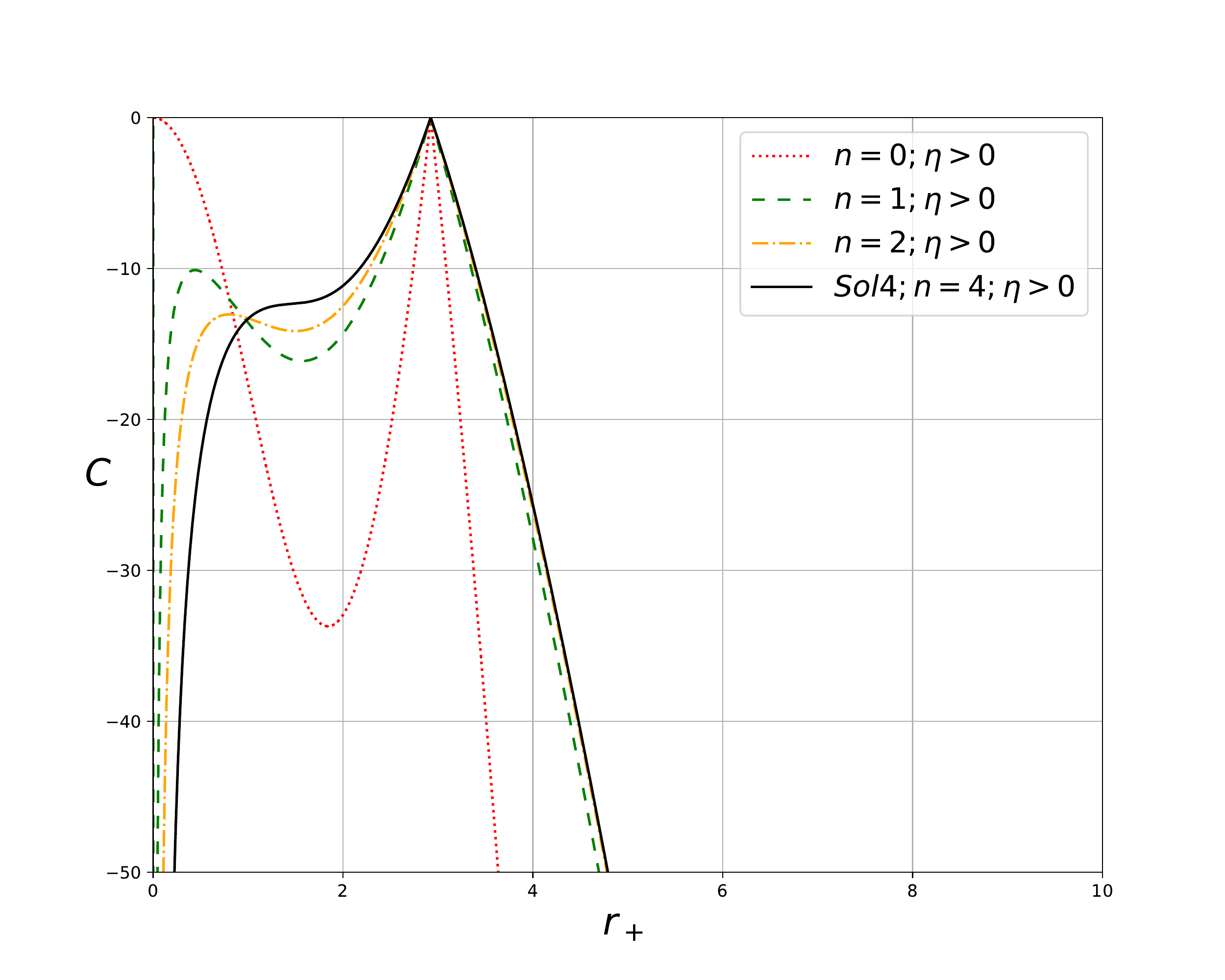}
         \caption{$\eta = 0.9$}
     \end{subfigure}
     \hfill
     \begin{subfigure}[b]{0.45\textwidth}
         \centering
         \includegraphics[width=1.2\textwidth]{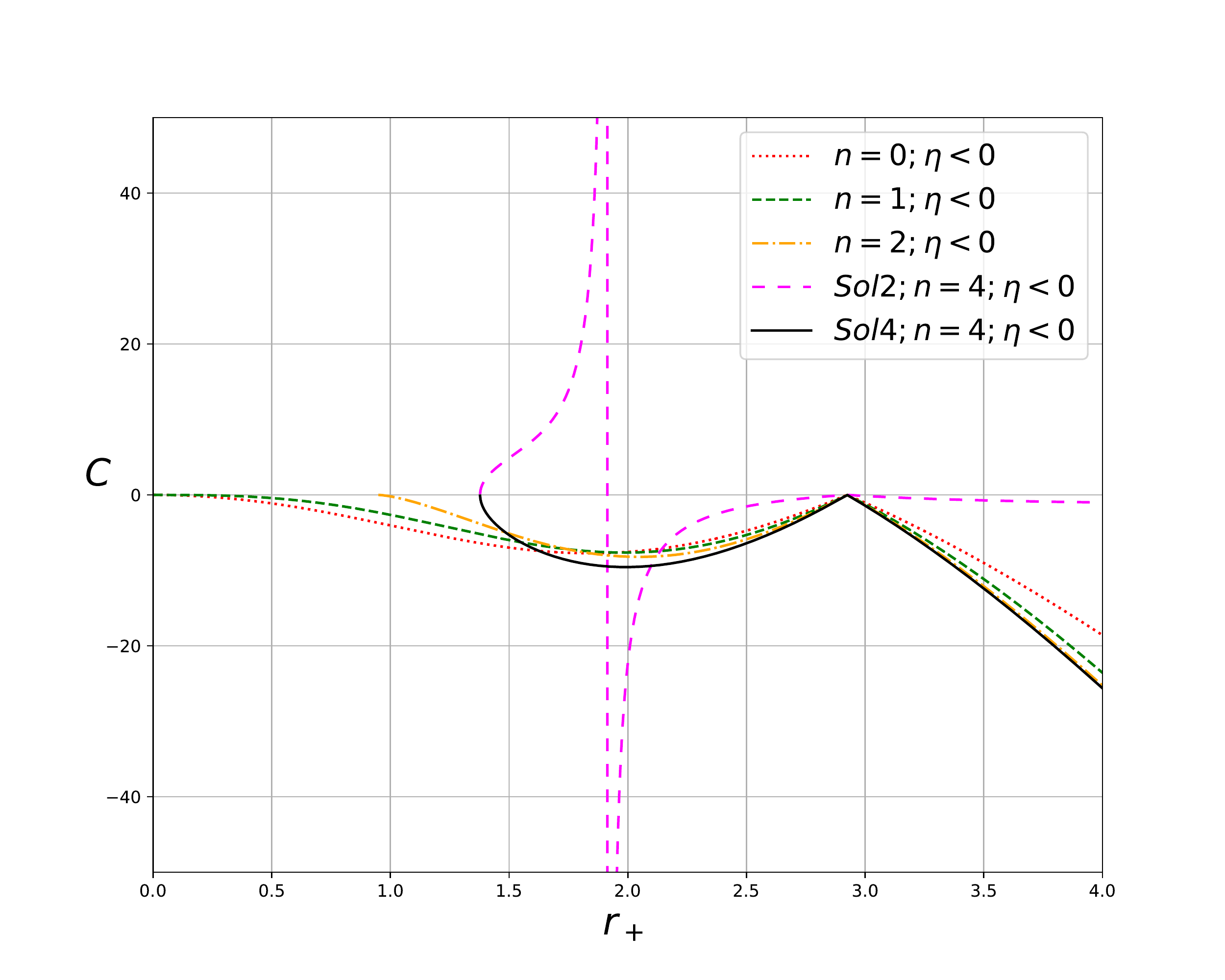}
         \caption{$\eta = -0.9$}
     \end{subfigure}
     \hfill
        \caption{Specific heat as function of the horizon radius, for Kiselev holes, in Rainbow gravity.}
        \label{fig:specificHeatKiselevRainbow}
\end{figure}

Now let's consider the complete case of Kiselev black hole in Rainbow, applied to the case of phantom dark energy. This is depicted in figure (\ref{fig:specificHeatKiselevRainbow}). What we perceive is a complementary behavior, the explanation of which is based on previous comments. 

Once again, the most interesting case occurs for $ n = 4 $, when we have an asymptotic behavior for specific heat. Note that this asymptotic behavior occurs far from the critical radius and it is due to the change that occurs in the relationship between temperature and horizon radius, which will allow the loop one can see in figure (\ref{fig:TempKiselevRainbow}). 

\subsection{Black Hole Remnants}

One of the most interesting subjects in black hole physics is the concept of remnants. What is a remnant and when can we be sure of its existence, however, does not seem to be a well-defined issue. The main problem is that often the solutions we seek are not valid for all the values of the horizon radius, so it is not clear whether what the study gives us is a remnant or just a problem in the coordinate system.

Black hole remnants in Rainbow has been studied in \cite{Mu:2015qna,Gim:2014ira, Ali:2014xqa}, and it has been claimed the existence of remnants. However, the remainder would occur at a non-zero finite temperature, which does not seem physically acceptable. A non-zero finite temperature black hole should continue to radiate. In this work, we will consider that remnants must be defined as objects with finite mass/radius whose temperature is zero or, in a weaker condition, approaches zero.

With this condition, we found no evidence that Rainbow gravity allows the existence of remnants by itself, at least for the studied Rainbow functions. The only exception occurs for $ n=4 $ with $ \eta<0 $, however the remainder could only be found in the presence of a cosmic fluid, and its final horizon radius is that where the event horizon and the cosmological horizon coincide. 

\section{Conclusions}

In this paper, we study thermodynamics and the possibility of  black hole remnants, immersed in a Kiselev metric, in Rainbow Gravity. We did a thorough and graphic analysis of several parameters of the theory, such as mass/internal energy, temperature, entropy and specific heat, both prioritizing Kiselev's metric and the Rainbow functions. 

We have explicitly shown that there is a limit value for the event horizon, in the case of Kiselev, which depends on the mass of the black hole. For this reason, there is a limit value for the temperature/entropy of this spacetime. It is for this reason that some authors find a negative temperature in their work, when they do not limit the values for the horizon radius. Such negative temperature is nothing more than an indication that beyond this critical horizon there is no mass that provides such event horizon, that is, it is actually a naked singularity.

When we consider Rainbow gravity, we have been able to find a finite value for temperature as the horizon radius decreases. This behavior does not occur for the Schwarzschild black hole, where the temperature tends to infinity as the horizon radius decreases. Although this case does not result in a remnant, at least it avoids the catastrophic behavior present in general relativity.

For a particular case of the MDR, where the energy is elevated to the fourth order, we find a new behavior in the thermodynamics of black holes, where a kind of bouncing occurs when we plot the temperature by the horizon radius. In this case, we also find a range in parameter space where the specific heat is positive, a behavior that is also unusual in general relativity. 

When we incorporate the Kiselev metric, in the latter case, a loop occurs in the temperature graph, which indicates the possibility of a remnant, albeit an unusual remnant, as it would occur for a horizon radius where the event horizon meets the cosmological horizon. With the exception of the latter case, we believe that Rainbow gravity with the proposed Rainbow functions is not a more open environment for the emergence of remnants than, for example, general relativity.

\acknowledgments
J.P.M.G. (grant 151701/2020-2) and V.B.B. would like to thank CNPq (Conselho Nacional de Desenvolvimento Cient\'ifico e Tecnol\'ogico - Brazil) for financial support. This study was financed in part by the Coordena\c{c}\~ao de Aperfei\c{c}oamento de Pessoal de N\'ivel Superior - Brasil (CAPES) - Finance Code 001. PHM would like to thanks CAPES for financial support.

\end{document}